# Coherent control of a structural phase transition in a solid-state surface system

Jan Gerrit Horstmann[1,*], Bareld Wit[1], Gero Storeck[1] & Claus Ropers[1,2,*]

[1] 4th Physical Institute, Solids and Nanostructures, University of Göttingen, Göttingen 37077, Germany.
[2] International Center for Advanced Studies of Energy Conversion (ICASEC), University of Göttingen, Göttingen 37077, Germany.

Correspondence to: jan-gerrit.horstmann@uni-goettingen.de (J.G.H.), cropers@gwdg.de (C.R.)

**The desire to exert active optical control over matter is a unifying theme across multiple scientific disciplines, as exemplified by all-optical magnetic switching[1,2], light-induced metastable or exotic phases of solids[3–9] and the coherent control of chemical reactions[10,11]. Typically, these approaches dynamically steer a system towards states or reaction products far from equilibrium. In solids, metal-insulator transitions are an important target for optical manipulation, offering dramatic and ultrafast changes of the electronic[4,5] and lattice[12–18] properties. In this context, essential questions concern the role of coherence in the efficiencies and thresholds of such transitions. Here, we demonstrate coherent vibrational control over a metal-insulator structural phase transition in a quasi-one-dimensional solid-state surface system. An optical double-pulse excitation scheme[19–22] is used to drive the system from the insulating to a metastable metallic state, and the corresponding structural changes are monitored by ultrafast low-energy electron diffraction[23–25]. We observe strong oscillations in the switching efficiency as a function of the double-pulse delay, revealing the importance of vibrational coherence in two key structural modes governing the transition on a femtosecond timescale. This mode-selective coherent control of solids and surfaces could open new routes to switching chemical and physical functionalities, facilitated by metastable and non-equilibrium states.**

Femtochemistry entails the search for understanding and control of ultrafast reaction pathways[10,22]. To this end, coherences in the electronic and vibrational states of reactants are employed to guide the system across a complex, generally multidimensional energy landscape[10,26]. Established for small molecules, a possible transfer of this concept to extended systems and solids is complicated, e.g. due to a high electronic and vibrational density of states, and couplings to an external heat bath[27]. Low-dimensional and strongly correlated systems represent a promising intermediate between molecules and solids, with phase transitions assuming the role of a "reaction". A number of these transitions can be driven optically - either by means of transient heating[24,28], electronic excitation[15–18,29,30] or direct resonant coupling to certain vibrational degrees of freedom[4–7,31]. The prototypical case of a phase transition governed by structural modes is given by the Peierls instability[32,33], in which a metal-to-insulator transition (MIT) is linked to phonon softening and the appearance of a static periodic lattice distortion (PLD). Coherent oscillations of the PLD, so-called amplitude modes or amplitudons, are frequently observed in the optical pumping of such transitions, especially close to their threshold[34–39]. In analogy to the vibrational spectroscopy of reacting molecules[40], amplitudons can be used to track ultrafast changes in the lattice symmetry across a phase transition[35,38,39]. However, it remains to be shown that a sequential excitation of coherent amplitude modes can be harnessed to drive a system non-thermally across a phase transition barrier, as proposed by Nelson, Weiner, *et al.* almost 30 years ago[20,41].

Here, we report coherent control over the phase transition in a quasi-one-dimensional Peierls insulator by manipulating the amplitudes of decisive phonon modes. We employ a double-pulse excitation scheme and monitor the structural transformation by ultrafast low-energy electron diffraction (ULEED; Fig. 1a, see Methods)[10–12]. Observing the resulting structure as a function of the double-pulse separation demonstrates the importance and distinct roles of two phonon modes on the femtosecond timescale.

As a model system, we study atomic indium wires on the Si(111) surface[42,43], a prominent Peierls system attracting significant interest for its ultrafast dynamics[15–18]. Arranged in a "zigzag" pattern, the indium atoms induce a metallic (4×1) superstructure, which, at $T_c$ = 125 K, exhibits a first-order transition to an insulating state with quadrupled (8×2) unit cell size and a "hexagon"-shaped indium pattern[44]. The associated change in atomic structure causes additional spots in backscattering diffraction (cf. LEED patterns in Fig. 1b). Below $T_c$, a single optical pump pulse is able to electronically excite the system to a metastable (4×1) state[15–18]. Recently, time-resolved diffraction and photoemission spectroscopy revealed the ultrafast nature of this transition (occuring on a 350 fs timescale) and identified excited electrons and localized photoholes as its driving force[16–18].

Tracking the (4×1)/(8×2) diffraction spot intensities in ULEED, we observe a rapid increase/decrease directly after optical excitation and subsequent relaxation to a level persisting over nanoseconds (Fig. 1c, left), evidencing the metastability of the structure[15,44]. Interestingly, this long-lived contribution displays a rather gradual threshold in pump fluence. This implies that for intermediate excitation densities, a variable part of the surface is switched to the metastable state (Fig. 1c, right), despite a homogeneous excitation of the probed area (see Methods). An interpretation based on the coexistence of both phases is also corroborated by scanning tunneling microscopy[45] and Raman spectroscopy[46] well below $T_c$.

It may be anticipated that near the threshold, the structural transition is particularly susceptible to weak perturbations, affecting the efficiency of driving the system to the metastable state. Motivated by control schemes in femtochemistry[19–22], we explore the use of pulse sequences to manipulate the switching efficiency. Specifically, we employ a pair of optical pump pulses with variable delay $\Delta t_{p-p}$, and probe the resulting structure by ULEED at a later time of $\Delta t_{p-el}$ = 75 ps, well after the excitation (Fig. 2a). We find that the signature of the metastable state, i.e., a mirror-like suppression/increase of the (8×2)/(4×1) phase, is a strong function of the pump-pump delay $\Delta t_{p-p}$. In particular, we identify a maximum transition yield for very small inter-pulse delays (Fig. 2b). Moreover, pronounced oscillations with a period of 1-2 ps are observed on either delay side. While we attribute the peaked signal around $\Delta t_{p-p}$ = 0 to additive electronic excitation and its decay on a few-picosecond timescale[47], the oscillations evidence a coherent response of the transition to the excitation. The frequency range of 0.5-1 THz points to a vibrational origin[46], which we further investigate below.

To examine the leverage of this double-pulse excitation, we measure the delay-dependent switching efficiency for a range of pump fluences (Fig. 2c). Whereas only a minor delay-dependence is found well below and above threshold, the measurements exhibit a strong modulation at intermediate fluences between 0.5 and 1.0 mJ cm$^{-2}$. We discuss these observations in light of the established potential energy model of the transition[15,16,48]. The (8×2)→(4×1) transformation is typically described in terms of a tristable energy surface, with an initial minimum at the (8×2) configuration and electronic excitation continuously tilting the balance towards the (4×1) phase (Fig. 2d)[15,43,47]. The presence of signal oscillations only at intermediate fluences suggests that vibrational motion contributes to overcoming a sufficiently lowered but not completely vanishing barrier, as illustrated in Fig. 2e. In this scenario, the first pump pulse leaves the system oscillating around an elevated (8×2) minimum, triggered by a displacive excitation of coherent phonons (DECP)[29]. For the reduced barrier, the timing of the second pump with respect to the oscillation of the phonon wavepacket plays a critical role: In-phase excitation results in a further increase of the vibrational energy and a barrier-crossing to the (4×1) state (① in Fig. 2e). Anti-phase excitation, on the other hand, vibrationally de-excites the system, which then has insufficient kinetic energy and remains in the (8×2) state (② in Fig. 2e). This illustration indicates a sharp threshold and a binary reaction outcome. The fraction of the sample surface switched to the metastable state is, however, a continuous function of the control variable $\Delta t_{p-p}$, due to variations in barrier height and local environment which also cause the smooth threshold in Fig. 1c.

In order to gain deeper insight into this control mechanism and to identify the structural modes governing the phase transition, we analyze the oscillatory response in greater detail. Figure 3b presents a trace of the (8×2) suppression at medium fluence, and its Fourier transform clearly shows two peaks at 0.57 THz and 0.83 THz (see inset). Importantly, these frequencies correspond to those of a shear mode ($f_s$ = 0.54 THz) and a rotation mode ($f_r$ = 0.81 THz) of the (8×2) structure (Fig. 3g), which were previously linked to the structural transition[48,49]. Specifically, a linear combination of the eigenvectors of the two modes connects both phases[50]. A short-time Fourier transform (Fig. 3c) further reveals delay-dependent frequency shifts of both modes, in particular a softening of the shear mode and a hardening of the rotation mode towards time-zero. Moreover, in two-pulse experiments with unequal pulse energies (Fig. 3d-f), coherences are more pronounced if the weaker pump pulse arrives first. The absence of signal oscillations for a strong first pulse likely originates from that pulse already switching a large fraction of the sample surface.

The distinct roles of the shear and rotation modes for the phase transition efficiency are evident from Fourier-filtered traces of the modulated switching efficiency (Fig. 3h). We first remove the quasi-DC component (solid black curve), which decays on a 3-ps time scale and corresponds to electronic relaxation[47]. The remaining oscillatory components are plotted individually (gold, pink) and as a sum (violet). With a maximum positive contribution at $\Delta t_{p-p}$ = 0, the shear-mode behaves as expected, i.e., in-phase excitation promotes the transition. Surprisingly, however, the rotation mode has an opposite effect, reducing the phase transition efficiency at zero pump-pump delay. The sum of both bands thus shows a small dip in switching efficiency at $\Delta t_{p-p} = 0$, which is also evident from the raw data (Fig. 2b) and was consistently observed in multiple measurements.

These counterintuitive observations call for a description in terms of a two-dimensional potential energy surface (PES) spanned by the rotation and shear mode displacements. Most importantly, the relevance of the rotation mode for the transition[16,49] needs to be reconciled with its apparent negative role for the switching efficiency. In a reasonable assumption for a model PES, optical pump pulses induce a displacive excitation of both modes along the line connecting the (8×2) and (4×1) states. Our experimental findings now suggest a scenario with a transition state offset from this line (Fig. 4a), specifically involving a large displacement of the shear and a much smaller displacement of the rotation mode. In this way, the transition state can be reached more efficiently for a de-excitation of the rotation mode. Figure 4b illustrates trajectories in such a simplified model PES, exhibiting a saddle-point transition state along the shear axis (see Methods for details). Top, middle and bottom planes show the potential and trajectories before, in-between and after both pump pulses, respectively. Three exemplary trajectories at varying pump-pump delay are drawn. For pump-pump overlap, the transition involves large rotational displacements (dark red). A small delay of the second pump partially de-excites the faster rotation mode, whereas the excitation of the slower shear mode remains additive (light red). In conjunction, enhanced shear-mode and suppressed rotation-mode excitation most efficiently overcomes the rotational bottleneck at the transition state. Finally, increasing the pump-pump delay to half a shear mode period (violet) hinders the transition entirely, leaving the system in a rotationally excited state.

We note that the experimentally observed softening of the shear mode near pump-pump overlap is also consistent with its role as the primary reaction coordinate at the saddle point. The hardening of the rotation mode, on the other hand, may indicate a respective narrowing of the potential energy surface near the transition state. However, further experimental and theoretical studies, involving density functional theory and molecular dynamics simulations, may elucidate the PES and the influence of nonlinear mode-couplings in detail. Finally, the microscopic excitation mechanism underlying the phonon coherences deserves further consideration, including its link to the femtosecond electron transfer and hole-induced driving forces recently described[17].

In conclusion, we have demonstrated the coherent control of a surface structural phase transition by all-optical manipulation of key phonon modes. Our results show that the outcome of the phase transition, much like many chemical reactions, depends on the coherent vibrational amplitude, highlighting the ballistic motion of the order parameter in overcoming the barrier. In molecular chemistry, it has long been known that vibrational excitation may drastically affect reaction rates, a fundamental principle captured by the Polanyi rules[26]. Our work extends this principle to surfaces and solids and introduces the vibrational phase as a decisive parameter to target the transition state. Moreover, we believe that exploiting vibrational coherences in low-dimensional and strongly correlated materials, as well as molecular adsorbates, holds promise for structural and electronic control in surface physics and chemistry, providing a handle to steer physical functionality and chemical reactivity.

1. Kimel, A. V. *et al.* Ultrafast non-thermal control of magnetization by instantaneous photomagnetic pulses. *Nature* **435**, 655–657 (2005).
2. Schlauderer, S. *et al.* Temporal and spectral fingerprints of ultrafast all-coherent spin switching. *Nature* **569**, 383 (2019).
3. Stojchevska, L. *et al.* Ultrafast switching to a stable hidden quantum state in an electronic crystal. *Science* **344**, 177–180 (2014).
4. Rini, M. *et al.* Control of the electronic phase of a manganite by mode-selective vibrational excitation. *Nature* **449**, 72–74 (2007).
5. Tobey, R. I., Prabhakaran, D., Boothroyd, A. T. & Cavalleri, A. Ultrafast electronic phase transition in $La_{1/2}Sr_{3/2}MnO_4$ by coherent vibrational excitation: Evidence for nonthermal melting of orbital order. *Phys. Rev. Lett.* **101**, 197404 (2008).
6. Mitrano, M. *et al.* Possible light-induced superconductivity in $K_3C_{60}$ at high temperature. *Nature* **530**, 461–464 (2016).
7. Nova, T., Disa, A., Fechner, M. & Cavalleri, A. Metastable ferroelectricity in optically strained $SrTiO_3$. *ArXiv181210560 Cond-Mat* (2018).
8. Sie, E. J. *et al.* An ultrafast symmetry switch in a Weyl semimetal. *Nature* **565**, 61 (2019).
9. Wang, Y. H., Steinberg, H., Jarillo-Herrero, P. & Gedik, N. Observation of Floquet-Bloch states on the surface of a topological insulator. *Science* **342**, 453–457 (2013).
10. Zewail, A. H. Femtochemistry: Atomic-scale dynamics of the chemical bond using ultrafast lasers (Nobel lecture). *Angew. Chem. Int. Ed.* **39**, 2586–2631 (2000).
11. Nuernberger, P., Vogt, G., Brixner, T. & Gerber, G. Femtosecond quantum control of molecular dynamics in the condensed phase. *Phys. Chem. Chem. Phys.* **9**, 2470–2497 (2007).
12. Cavalleri, A. *et al.* Femtosecond structural dynamics in $VO_2$ during an ultrafast solid-solid phase transition. *Phys. Rev. Lett.* **87**, 237401 (2001).
13. Morrison, V. R. *et al.* A photoinduced metal-like phase of monoclinic $VO_2$ revealed by ultrafast electron diffraction. *Science* **346**, 445–448 (2014).
14. Liu, M. *et al.* Terahertz-field-induced insulator-to-metal transition in vanadium dioxide metamaterial. *Nature* **487**, 345–348 (2012).
15. Wall, S. *et al.* Atomistic picture of charge density wave formation at surfaces. *Phys. Rev. Lett.* **109**, 186101 (2012).
16. Frigge, T. *et al.* Optically excited structural transition in atomic wires on surfaces at the quantum limit. *Nature* **544**, 207–211 (2017).
17. Nicholson, C. W. *et al.* Beyond the molecular movie: Dynamics of bands and bonds during a photoinduced phase transition. *Science* **362**, 821–825 (2018).
18. Chávez-Cervantes, M., Krause, R., Aeschlimann, S. & Gierz, I. Band structure dynamics in indium wires. *Phys. Rev. B* **97**, 201401 (2018).
19. Hase, M., Fons, P., Mitrofanov, K., Kolobov, A. V. & Tominaga, J. Femtosecond structural transformation of phase-change materials far from equilibrium monitored by coherent phonons. *Nat. Commun.* **6**, 8367 (2015).
20. Weiner, A. M., Leaird, D. E., Wiederrecht, G. P. & Nelson, K. A. Femtosecond pulse sequences used for optical manipulation of molecular motion. *Science* **247**, 1317–1319 (1990).

21. Feurer, T., Vaughan, J. C. & Nelson, K. A. Spatiotemporal coherent control of lattice vibrational waves. *Science* **299**, 374–377 (2003).
22. Potter, E. D., Herek, J. L., Pedersen, S., Liu, Q. & Zewail, A. H. Femtosecond laser control of a chemical reaction. *Nature* **355**, 66 (1992).
23. Gulde, M. *et al.* Ultrafast low-energy electron diffraction in transmission resolves polymer/graphene superstructure dynamics. *Science* **345**, 200–204 (2014).
24. Vogelgesang, S. *et al.* Phase ordering of charge density waves traced by ultrafast low-energy electron diffraction. *Nat. Phys.* **14**, 184–190 (2018).
25. Storeck, G., Vogelgesang, S., Sivis, M., Schäfer, S. & Ropers, C. Nanotip-based photoelectron microgun for ultrafast LEED. *Struct. Dyn.* **4**, 044024 (2017).
26. Polanyi, J. C. & Wong, W. H. Location of energy barriers. I. Effect on the dynamics of reactions A + BC. *J. Chem. Phys.* **51**, 1439–1450 (1969).
27. Cavalleri, A. *et al.* Anharmonic lattice dynamics in germanium measured with ultrafast X-ray diffraction. *Phys. Rev. Lett.* **85**, 586–589 (2000).
28. Haupt, K. *et al.* Ultrafast metamorphosis of a complex charge-density wave. *Phys. Rev. Lett.* **116**, 016402 (2016).
29. Zeiger, H. J. *et al.* Theory for displacive excitation of coherent phonons. *Phys. Rev. B* **45**, 768–778 (1992).
30. Sciaini, G. *et al.* Electronic acceleration of atomic motions and disordering in bismuth. *Nature* **458**, 56–59 (2009).
31. Fausti, D. *et al.* Light-induced superconductivity in a stripe-ordered cuprate. *Science* **331**, 189–191 (2011).
32. Peierls, R. E. *Quantum theory of solids*. (Oxford University Press, 2001).
33. Fröhlich Herbert. On the theory of superconductivity: the one-dimensional case. *Proc. R. Soc. Lond. Ser. Math. Phys. Sci.* **223**, 296–305 (1954).
34. Eichberger, M. *et al.* Snapshots of cooperative atomic motions in the optical suppression of charge density waves. *Nature* **468**, 799–802 (2010).
35. Wall, S. *et al.* Ultrafast changes in lattice symmetry probed by coherent phonons. *Nat. Commun.* **3**, 721 (2012).
36. Sokolowski-Tinten, K. *et al.* Femtosecond X-ray measurement of coherent lattice vibrations near the Lindemann stability limit. *Nature* **422**, 287–289 (2003).
37. Rettig, L., Chu, J.-H., Fisher, I. R., Bovensiepen, U. & Wolf, M. Coherent dynamics of the charge density wave gap in tritellurides. *Faraday Discuss.* **171**, 299–310 (2014).
38. Beaud, P. *et al.* A time-dependent order parameter for ultrafast photoinduced phase transitions. *Nat. Mater.* **13**, 923–927 (2014).
39. Neugebauer, M. J. *et al.* Optical control of vibrational coherence triggered by an ultrafast phase transition. *ArXiv190200388 Cond-Mat* (2019).
40. Nibbering, E. T. J., Fidder, H. & Pines, E. Ultrafast Chemistry: Using time-resolved vibrational spectroscopy for interrogation of structural dynamics. *Annu. Rev. Phys. Chem.* **56**, 337–367 (2005).
41. Nelson, K. A. The prospects for impulsively driven, mode-selective chemistry in condensed phases. in *Mode Selective Chemistry* (eds. Jortner, J., Levine, R. D. & Pullman, B.) 527–533 (Springer Netherlands, 1991).

42. Yeom, H. W. *et al*. Instability and charge density wave of metallic quantum chains on a silicon surface. *Phys. Rev. Lett*. **82**, 4898–4901 (1999).
43. Snijders, P. C. & Weitering, H. H. Colloquium: Electronic instabilities in self-assembled atom wires. *Rev. Mod. Phys*. **82**, 307–329 (2010).
44. Klasing, F. *et al*. Hysteresis proves that the In/Si(111) (8×2) to (4×1) phase transition is first-order. *Phys. Rev. B* **89**, 121107 (2014).
45. Terada, Y. *et al*. Optical doping: active control of metal–insulator transition in nanowire. *Nano Lett*. **8**, 3577–3581 (2008).
46. Speiser, E., Esser, N., Wippermann, S. & Schmidt, W. G. Surface vibrational Raman modes of In:Si(111) (4×1) and (8×2) nanowires. *Phys. Rev. B* **94**, 075417 (2016).
47. Nicholson, C. W. *et al*. Excited-state band mapping and momentum-resolved ultrafast population dynamics in In/Si(111) nanowires investigated with XUV-based time- and angle-resolved photoemission spectroscopy. *Phys. Rev. B* **99**, 155107 (2019).
48. Jeckelmann, E., Sanna, S., Schmidt, W. G., Speiser, E. & Esser, N. Grand canonical Peierls transition in In/Si(111). *Phys. Rev. B* **93**, 241407 (2016).
49. Wippermann, S. & Schmidt, W. G. Entropy explains metal-insulator transition of the Si(111)-In nanowire array. *Phys. Rev. Lett*. **105**, 126102 (2010).
50. Schmidt, W. G. *et al*. In-Si(111)(4×1)/(8×2) nanowires: Electron transport, entropy, and metal-insulator transition. *Phys. Status Solidi B* **249**, 343–359 (2012).

**Acknowledgements** This work was funded by the European Research Council (ERC Starting Grant 'ULEED', ID: 639119) and the Deutsche Forschungsgemeinschaft (SFB-1073, project A05). We gratefully acknowledge insightful discussions with H. Böckmann, F. Kurtz, S. Vogelgesang, H. Schwoerer, T. Elsässer, M. Wolf and M. Horn-von Hoegen.

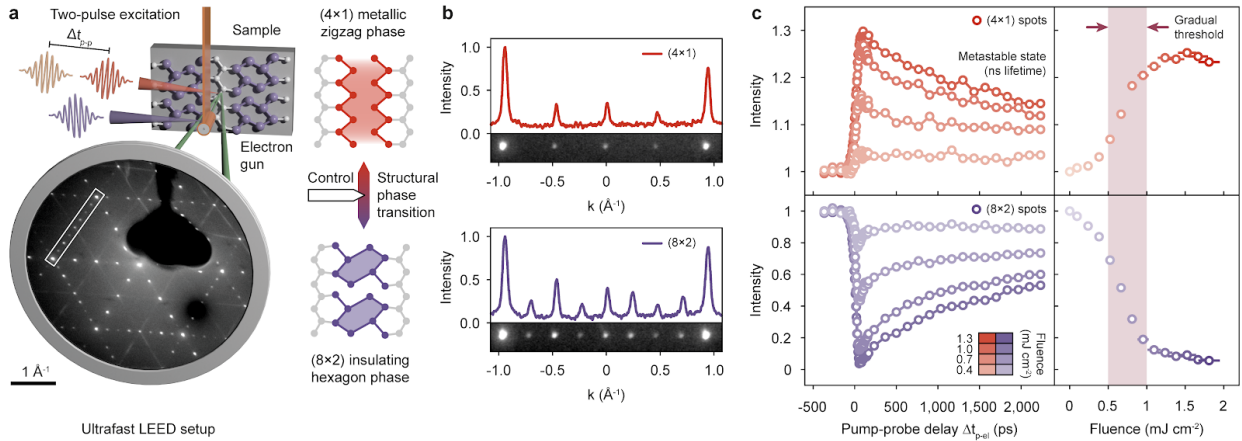

**Figure 1 | Ultrafast LEED setup and structural phase transition in atomic indium wires on silicon.**

**a,** Experimental scheme. Ultrashort electron pulses from a miniaturized laser-driven electron gun are utilized in a LEED experiment to monitor the microscopic structure of atomic indium wires on the Si(111) surface after optical excitation with single or double pulses. **b,** Cutouts and line profiles from LEED patterns of the metallic (4×1) and insulating (8×2) phases (white frame in a). The emergence of additional spots in the (8×2) phase indicates the pronounced structural changes during the phase transition. **c,** (Left) Time-resolved integrated intensities of (4×1) and (8×2) diffraction spots as a function of the pump-probe delay $\Delta t_{p\text{-}el}$. (Right) Fluence-dependent spot intensities recorded at $\Delta t_{p\text{-}el} = 75$ ps. The (4×1) and (8×2) intensities have been normalized to corresponding values at $I(\Delta t_{p\text{-}el} < 0)$.

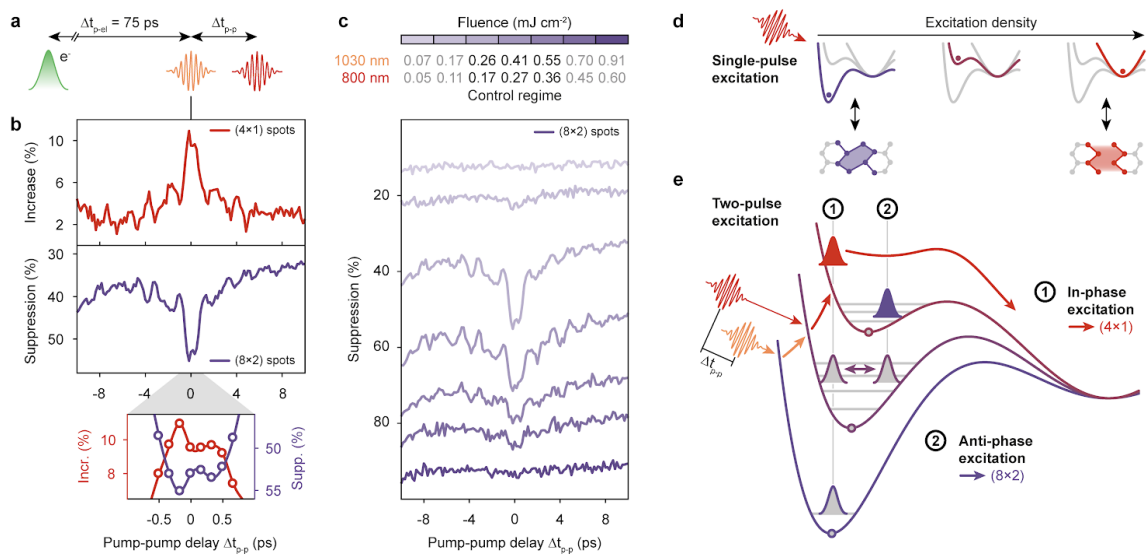

**Figure 2 | Coherent control of the (8×2)→(4×1) phase transition.**
**a**, Sketch of the pulse sequence in pump-pump-probe experiments. **b**, Increase/suppression of the integrated (4×1)/(8×2) diffraction spot intensity as a function of the pump-pump delay $\Delta t_{p\text{-}p}$, revealing oscillations of the phase transition efficiency (incident fluences of the 1030 nm and 800 nm pulses: $F_{1030}$ = 0.26 mJ cm$^{-2}$; $F_{800}$ = 0.17 mJ cm$^{-2}$). **c**, Delay-dependent suppression of the (8×2) diffraction spots for different combined fluences up to 1.51 mJ cm$^{-2}$. Pronounced oscillations are only observed between 0.5 and 1.0 mJ cm$^{-2}$. **d**, Phase transition model based on reshaping of the tristable energy surface by a single pump pulse. For simplicity, the second, energetically degenerate (8×2) minimum is not depicted. Note that the potential deformation is a continuous function of the excitation density. **e**, Illustration of the importance of vibrational coherence for the transition within the intermediate fluence regime in double-pulse experiments (see text).

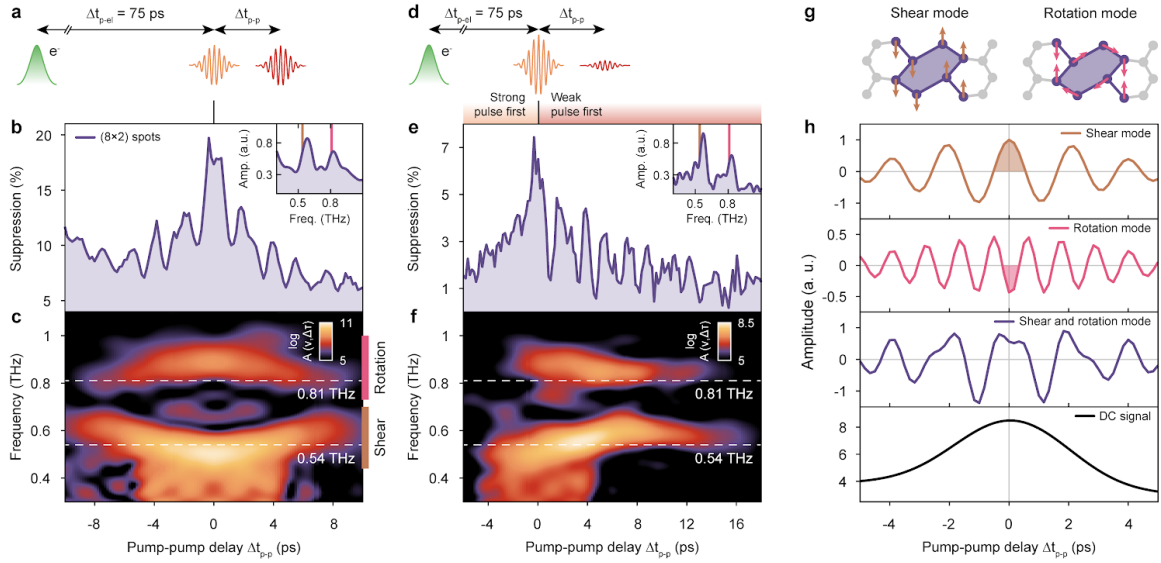

**Figure 3 | Fourier analysis of the phase transition efficiency in double-pump experiments.**
**a**, **d**, Sketches of pump-pump-probe experiments with equal (a) and unequal (d) pulses. **b**, **e**, Suppression of the (8×2) diffraction spots as a function of the pump-pump delay $\Delta t_{p-p}$ for equal (b) and unequal (e) pulses. Insets: Fourier transforms of the oscillations in (b) and (e) with reference frequencies of (8×2) shear (0.54 THz, orange line) and rotation (0.81 THz, pink line) modes (a.u.: arbitrary units). **c, f**, Short-time Fourier transforms of the data in (b) and (e)**.** For comparison, dashed lines indicate the reference frequencies of shear and rotation modes. **g**, Eigenvectors of the (8×2) shear and rotation modes governing the structural transition into the (4×1) phase. **h**, Frequency-specific contributions to the transition efficiency as a function of $\Delta t_{p-p}$ obtained by Fourier filtering the signal plotted in (b): Shear mode (gold), rotation mode (pink), shear and rotation mode (violet), quasi-DC background (black). Amplitudes have been normalized to the shear mode contribution at $\Delta t_{p-p} = 0$.

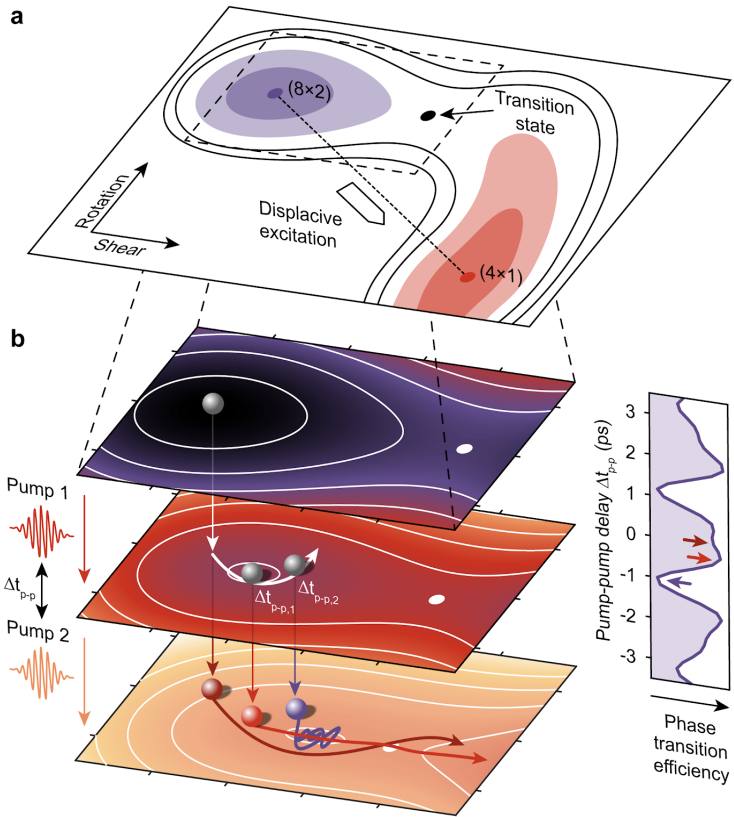

**Figure 4 | Two-dimensional picture of the phase transition dynamics.**
**a**, Proposed two-dimensional model of the potential energy surface (PES) for the (8×2)→(4×1) transition. In the vicinity of the (8×2) state, the eigenfrequencies of the PES correspond to shear and rotation mode frequencies. Displacive excitation occurs along the connecting line between (8×2) and (4×1) states, i.e. at an angle to the transition state located along the shear axis. **b**, Sketch of exemplary system trajectories close to the (8×2) state before (top), in between (middle) and after (bottom) two subsequent displacive excitations of the PES. The pump-pump delay $\Delta t_{p-p}$ controls the contributions of shear and rotational motion to the trajectories and thus the transition efficiency (side panel).

# Supplementary Information

**METHODS**

**Ultrafast LEED and experimental setup.** We recently developed ultrafast low-energy electron diffraction (ULEED) in an optical-pump/electron probe scheme for the time-resolved investigation of structural dynamics at solid state surfaces[1–3]. LEED is a surface-sensitive technique, in which the diffraction pattern of electrons backscattered from a sample is analysed to obtain information about the surface structure[4].

In order to achieve high temporal and momentum resolution, we use a laser-driven electron gun consisting of a nanometric tungsten tip as well as four metal electrodes (2 mm outer diameter, aperture diameter 400 µm), which act as a suppressor-extractor unit and an electrostatic einzel lens[2]. Electron pulses are generated via localized two-photon photoemission by illuminating the tip apex with femtosecond laser pulses (central wavelength 400 nm, pulse duration 45 fs, 20 nJ pulse energy) at repetition rates up to 100 kHz. The needle cathode provides a reduced electron beam emittance allowing for a momentum-resolution in diffraction of 0.03 Å$^{-1}$. Moreover, we lower the dispersion-induced electron pulse broadening effect by decreasing the propagation length between the electron source and the sample. In this respect, the reduced dimensions of the electron gun allow for operational distances of a few millimeters at a reasonably small fraction of shadowed electron diffraction signal, resulting in electron pulse durations down to 16 ps at the sample[2]. The backscattered electrons from the surface are amplified and recorded by a combination of a chevron micro-channel plate (MCP), a phosphor screen and a cooled sCMOS camera resulting in typical integration times of $t_{int}$ = 20 s per frame in time-resolved measurements.

In the pump-probe experiments (Fig. 1c), the surface structure is excited by ultrashort light pulses ($\lambda_c$ = 1030 nm, $E_p$ = 1.2 eV, $\Delta\tau$ = 212 fs) and probed by electron pulses ($E_{kin}$ = 80 eV) at a variable time delay $\Delta t_{p\text{-el}}$. To ensure a homogeneous excitation of the area probed by the electrons, we expand the optical pump beam to (297±13)×(223±14) µm² in the sample plane, which is significantly larger than the focal spot size of the electron gun (< 80×80 µm²).

For the coherent control of the structural phase transition between the (4×1) and the (8×2) phase (Fig. 2b,c; Fig. 3b,e), we use two pump pulses with distinct central wavelengths ($P_1$: $\lambda_c$ = 1030 nm, $E_p$ = 1.2 eV, $\Delta\tau$ = 212 fs; $P_2$: $\lambda_c$ = 800 nm, $E_p$ = 1.55 eV, $\Delta\tau$ = 232 fs) from a Yb:YAG amplifier system and an optical parametric amplifier (OPA) to avoid interference effects around time-zero (coherent artifacts). The $P_1$ and $P_2$ beams are aligned collinearly and subsequently focused onto the sample by a single lens (Fig. S1a). For finding the temporal overlap of the pump pulses, we perform cross-correlation measurements using a fast nonlinear photodiode (GaP) (Fig. S1b). A sketch of the experimental setup is depicted in Fig. S1a.

**Sample preparation.** All experiments were carried out under ultra-high vacuum conditions (base pressure p < 2×10$^{-10}$ mbar) in order to minimize surface defects from adsorption, which were found to have a significant influence on the formation of the low-temperature (8×2) phase as well and the lifetime of the metastable state[5,6]. The samples were prepared by flash-annealing

Si(111) wafers (phosphorous doped, resistivity R = 0.6-2 Ω cm) at $T_{max}$ = 1250 °C via direct current heating (maximum pressure during flashing was kept below $p_{max}$ = 2×10$^{-9}$ mbar). Evaporation of 1.2 monolayers of indium onto the resulting Si(111)(7×7) surface reconstruction at room temperature followed by subsequent annealing at T = 500 °C for 300 s resulted in a high-quality (4×1) phase, as verified in our ultrafast LEED setup. After inspection of the (4×1) phase, the samples were immediately cooled down to a base temperature of T = 60 K using an integrated continuous flow helium cryostat. The phase transition between the high-temperature (4×1) and the low-temperature (8×2) phase was observed at 125 K. LEED images of the (7×7), the (4×1) and the (8×2) structure are shown in Fig. S2.

**Data analysis.** The LEED pattern of the (8×2) phase from Fig. 1a and the cutouts shown in Fig. 1b were recorded at a base temperature of T = 60 K (cutout of the (4×1) phase: T = 300 K) with an integration time of $t_{int}$ = 60 s. The diffraction images are plotted on a logarithmic color scale to enhance the visibility of the twofold streaks, which are typically one order of magnitude weaker than the (8×2) spots. The location of the cutout regions within the complete diffraction image is indicated by the white rectangle in Fig. 1a.

For the analysis of both the pump-probe and the pump-pump-probe experiments, we sum up the background-corrected raw data peak intensities within circular areas of interest (radius $r$) around the selected (4×1) and (8×2) spots. To this end, the background is determined within a ring (width d$r$) around the edge of each area of interest. We use radii of $r$ = 0.10 Å$^{-1}$ (40 pixels) for the fluence-dependent data presented in Fig .2b, $r$ = 0.08 Å$^{-1}$ (30 pixels) for the data presented in Figs. 3b,e and a ring width of d$r$ = 0.008 Å$^{-1}$ (3 pixels) for all datasets.

To determine the relative changes in the (4×1) and (8×2) spot intensities caused by a single optical pulse (see Fig. 1c), the integrated peak intensities at the point of maximum suppression/enhancement ($\Delta t_{p-el}$ = 75 ps) are normalized to the value before time-zero.

The fluence-dependent suppression $s(\Delta t_{p-el}, \Delta t_{p-p}, F_{1030}, F_{800})$ of the (4×1) and (8×2) signals in the pump-pump-probe experiments (see Fig. 2b,d) is shown relative to the intensity $I(\Delta t_{p-el}$ = 75 ps, $F_{1030}$ = 0, $F_{800}$ = 0) without optical excitation. The curves shown in Figs. 3b,e are normalized to the intensities for a single pump pulse $I(\Delta t_{p-el}$ = 75 ps, $F_{1030}$ > 0, $F_{800}$ = 0) near maximum suppression.

**Fourier analysis.** In order to study the delay-dependent frequency change of both the shear and the rotation mode, we perform a short-time Fourier transform (STFT) of the datasets depicted in Figs. 3b and 3e with a super-gaussian window function

$$F_{filt,t} = exp(-(\frac{(t-t_{shift})^2}{2\sigma_t^2})^3) \quad (1)$$

in the time-domain, yielding the data shown in Figs. 3c and 3e ($\sigma_t$ = 3.6 ps). To extract the contributions of the individual modes to the signal from Fig. 3b, a super-gaussian frequency window

$$F_{filt,f} = exp(-(\frac{(f-f_c)^2}{2\sigma_f^2})^3) \quad (2)$$

is used to isolate the relevant frequency range. The data shown in Fig. 3h is obtained by an inverse FT of the filtered Fourier transform (shear mode: $f_c = 0.5$, $\sigma_f = 0.10$; rotation mode: $f_c = 0.9$, $\sigma_f = 0.07$; DC: $f_c = 0.0$, $\sigma_f = 0.14$).

**Two-dimensional model potential.** To discuss the roles of the shear and rotation modes for the phase transition efficiency (see Fig. 4), we compute exemplary trajectories in a simplified PES near the local (8×2) minimum (x = 0, y = 0) and the transition state. The underlying two-dimensional potential plotted in Fig. 4b (top) is given by

$$\Phi_0(x,y) = \tfrac{1}{2}(\omega_s^2 x^2 + \omega_r^2 y^2) - \tfrac{1}{2}a_3 x^3, \quad (3)$$

with $\omega_s/\omega_p = 0.54/0.81 = 2/3$. The two sequential displacive excitations of the PES are modeled via the time-dependent quadratic potential

$$\Phi_{pump}(x,y,\Delta\tau) = (p_x(x-x_p)^2 + p_y(y-y_p)^2) \cdot (\theta(0) + 2\theta(\Delta t_{p-p})) \quad (4)$$

centered around $(x_p, y_p)$. Here, $\Delta t_{p-p}$ is the pump-pump delay and $\theta$ the Heaviside function. To compute the trajectory $T(\Delta t_{p-p})$ of the system, we solve the classical equations of motion inside the potential $\Phi(x,y,\Delta\tau) = \Phi_0(x,y) + \Phi_{pump}(x,y,\Delta t_{p-p})$ using the ode45 solver (Dormand-Prince method) and the parameter set [$\omega_s = 1$, $\omega_r = 1.5$, $a_3 = 0.2$, $x_p = r_p \cdot \sin(\varphi_{disp})$, $y_p = r_p \cdot \cos(\varphi_{disp})$, $r_p = 10$, $\varphi_{disp} = 60°$]. We note that the goal of this model is not to predict quantitative transition rates, but rather to illustrate a scheme which accounts for the phase difference observed experimentally. Specifically, the physical PES will be non-separable in the vicinity of the saddle point, translating spatial and velocity deviations into changes of the transition rate. Moreover, for reasons of simplicity, only one of the four degenerate (8×2) ground states described by Cheon et al.[7] is considered in the simplified model potential illustrated in Fig. 4.

1. Gulde, M. *et al.* Ultrafast low-energy electron diffraction in transmission resolves polymer/graphene superstructure dynamics. *Science* **345**, 200–204 (2014).
2. Vogelgesang, S. *et al.* Phase ordering of charge density waves traced by ultrafast low-energy electron diffraction. *Nat. Phys.* **14**, 184–190 (2018).
3. Storeck, G., Vogelgesang, S., Sivis, M., Schäfer, S. & Ropers, C. Nanotip-based photoelectron microgun for ultrafast LEED. *Struct. Dyn.* **4**, 044024 (2017).
4. Van Hove, M. A., Weinberg, W. H. & Chan, C.-M. *Low-Energy Electron Diffraction: Experiment, Theory and Surface Structure Determination*. (Springer-Verlag, 1986).
5. Wall, S. *et al.* Atomistic picture of charge density wave formation at surfaces. *Phys. Rev. Lett.* **109**, 186101 (2012).
6. Klasing, F. *et al.* Hysteresis proves that the In/Si(111) (8×2) to (4×1) phase transition is first-order. *Phys. Rev. B* **89**, 121107 (2014).
7. Cheon, S., Kim, T.-H., Lee, S.-H. & Yeom, H. W. Chiral solitons in a coupled double Peierls chain. *Science* **350**, 182–185 (2015).

# SUPPLEMENTARY FIGURES

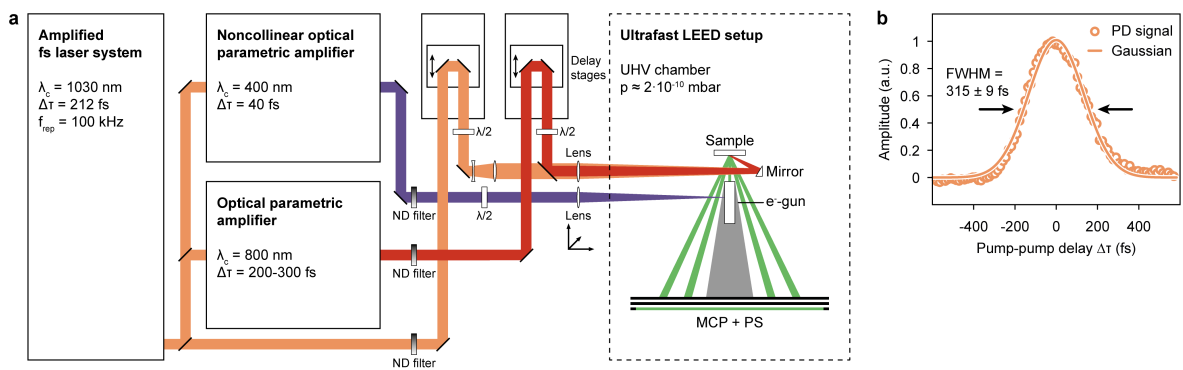

**Figure S1 | Experimental setup.**
**a**, Ultrashort laser pulses ($P_1$: $\lambda_c$ = 1030 nm, $\Delta\tau$ = 200 fs) from an Yb:YAG amplifier (left) pump a non-collinear OPA (output: $\lambda_c$ = 400 nm, $\Delta\tau$ = 40 fs) and an OPA (output: $P_2$, $\lambda_c$ = 800 nm, $\Delta\tau$ = 232 fs). The 1030 nm and 800 nm beams are independently attenuated and collinearly focused onto the sample by a single lens (400 mm focal length). The relative on-axis position of the two foci is controlled by adjusting the divergence of the 1030 nm beam. The UV pulses are focused onto the tungsten needle emitter inside the electron gun (e⁻-gun) to generate ultrashort electron pulses. The relative timing between the electron probe and each of the two optical pump pulses is controlled independently by two separate optical delay stages. The pump-induced changes in the LEED pattern are recorded using a micro-channel plate assembly. **b**, Cross-correlation of the two pump pulses recorded with a nonlinear photodiode to determine the temporal resolution of the pump-pump-probe experiment.

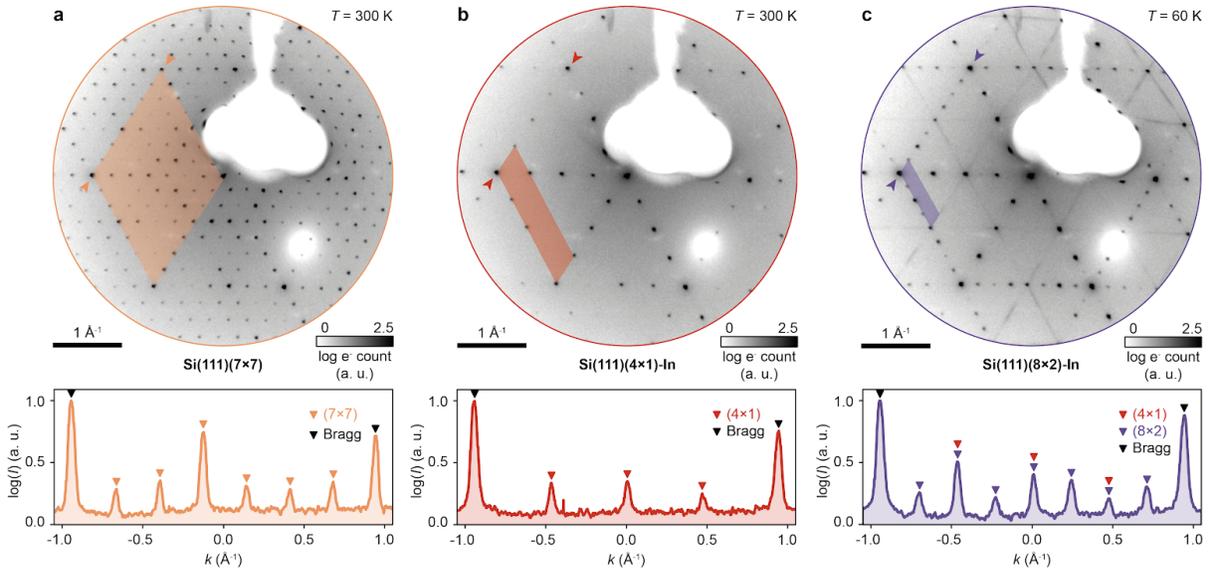

**Figure S2 | Diffraction images.**
Diffraction images and lineouts of the clean (7×7)-reconstructed Si(111) surface (a), the (4×1) (b) and (8×2) phase (c) recorded in our ultrafast LEED setup ($E_{kin}$ = 130 eV). Coloured areas correspond to the unit cells in reciprocal space, arrows indicate the location of the lineouts shown below. In the transformation from the (4×1) to the (8×2) phase, the unit cell is doubled in both dimensions. The two-fold streaks in the diffraction pattern of the (8×2) phase originate from a weak coupling between the atomic chains. The diffraction patterns of the indium-reconstructed phases feature contributions from three domains rotated by 120 degrees with respect to each other, since the hexagonal structure of the underlying substrate allows for three different orientations of the atomic indium chains.

16